\documentclass[english]{article}
\usepackage[T1]{fontenc}
\usepackage[latin9]{inputenc}
\usepackage[a4paper]{geometry}
\usepackage{amsmath}
\usepackage{setspace}
\usepackage{amssymb}
\makeatletter
\newcommand{\lyxaddress}[1]{
\par {\raggedright #1
\vspace{1.4em}
\noindent\par}
}

\makeatother

\usepackage{babel}

\begin{document}

\title{\textbf{Quaternion Gravi-Electromagnetism}}

\author{A. S. Rawat$^{\text{(1)}}$ and O. P. S. Negi $^{\text{(2)}}$}

\maketitle

\lyxaddress{\begin{center}
$^{\text{(1)}}$Department of Physics\\
 H. N. B. Garhwal University \\
Pauri Campus, \\
Pauri-246001 (U.K.), India
\par\end{center}}

\lyxaddress{\begin{center}
$^{\text{(2)}}$Department of Physics\\
 Kumaun University\\
 S.S.J.Campus\\
Almora- 263601 (U.K.), India
\par\end{center}}

\lyxaddress{\begin{center}
Email- drarunsinghrawat@gmail.com\\
ops\_negi@yahoo.co.in
\par\end{center}}
\begin{abstract}
Defining the generalized charge, potential, current and generalized
fields as complex quantities where real and imaginary parts represent
gravitation and electromagnetism respectively, corresponding field
equation, equation of motion and other quantum equations are derived
in manifestly covariant manner. It has been shown that the field equations
are invariant under Lorentz as well as duality transformations. It
has been shown that the quaternionic formulation presented here remains
invariant under quaternion transformations. 


\end{abstract}

\section{Introduction}

The idea of magnetic monopole was put forward by Dirac \cite{key-1}
in order to maintain the symmetry between electric and magnetic fields
in Maxwell's equations. The analogy between linear gravitational and
electromagnetic fields leads the asymmetry in Einstein's linear equation
of gravity and suggests the existence of gravitational analogue \cite{key-2}
of magnetic monopole \cite{key-3}. Like magnetic field, Cattani \cite{key-4}
introduced a new field (called Heavisidian field) depending upon the
velocities of gravitational charges (masses) \cite{key-2} and derived
the covariant equations (like Maxwell's equation) of linear gravitational
fields. On the other hand, some authors \cite{key-5} described the
existence of gravi-magnetic and gravi-electric fields instead of gravito-Heavisidian
fields \cite{key-6} associated with gravito-dyons (particle carrying
simultaneously gravitational and Heavisidian charges). Avoiding the
use of arbitrary string variables, Bisht et al \cite{key-7} has developed
manifestly covariant theory of gravito-dyons in terms of two four-potentials
and maintained the structural symmetry between generalized electromagnetic
field of dyons (particle carrying simultaneous existence of electric
and magnetic charges) \cite{key-8} and those of generalized gravito-Heavisidian
fields of gravito-dyons. Corresponding field equation and equation
of motion for unified fields of dyons and gravito-dyons are obtained
by Rajput \cite{key-6}. Quaternion formulation for generalized electromagnetic
fields of dyons and generalized gravito-Heavisidian fields of gravito-dyons
has also been developed \cite{key-9} in compact and consistent manner.
Accordingly, a consistent theory for the dynamics of four charges
(masses) (namely electric, magnetic, gravitational, Heavisidian) have
also been developed\cite{key-10} in simple, compact and consistent
manner. Considering an invariant Lagrangian density and its quaternionic
representation, the consistent field equations for the dynamics of
four charges have already been derived \cite{key-11} and it has been
shown that the present reformulation reproduces the dynamics of individual
charges (masses) in the absence of other charge (masses) as well as
the generalized theory of dyons (gravito - dyons) in the absence gravito
- dyons (dyons). Keeping in view the recent potential importance of
monopoles and the fact that the formalism necessary to develop them
has been clumsy and manifestly non-covariant as well as the recent
interest in linear gravity, in this paper, we have undertaken the
study of gravitation and electromagnetism together by defining the
complex four-potential, the real part of which represents gravitation
and the imaginary part describes electromagnetism. Defining the generalized
charge, potential, current and generalized fields as complex quantities
where real and imaginary parts represent the constituents of gravitation
and electromagnetism respectively, the generalized field equation
and equation of motion are obtained in consistent and manifestly covariant
manner. The suitable Lagrangian density for generalized fields of
gravitation and electromagnetic charges has been described to yield
the consistent form of corresponding field equation, equation of motion
and continuity equation in manifestly covariant way. The electric,
magnetic, gravitational and Heavisidian fields are discussed and Maxwell
like equations for linear gravity and electromagnetism are obtained
consistently in compact notation. It has been shown that the field
equations are invariant under quaternion Lorentz transformations and
duality transformations as well. The present theory reduces the the
theory of linear gravity (or electromagnetism) in the absence of electromagnetism
(gravitation) or vice versa.

\section{Generalized Gravi-electromagnetism}

Let us define the complex four-potential $\left\{ \mathsf{V}_{\mu}\right\} $
associated with gravi-electromagnetic field as

\begin{align}
\left\{ \mathsf{V}_{\mu}\right\} = & \left\{ \mathbf{\mathrm{B}}_{\mu}\right\} -\, i\,\left\{ \mathrm{A}_{\mu}\right\} \,\,\,\,\,\,\,\,\,\,\,(i=\sqrt{-1})\label{eq:1}\end{align}
where $\left\{ \mathbf{\mathrm{B}}_{\mu}\right\} $ and $\left\{ \mathrm{A}_{\mu}\right\} $
are respectively described as gravitational and electromagnetic four-potentials
for linear gravitational and electromagnetic fields. $\left\{ \mathbf{\mathrm{B}}_{\mu}\right\} $
and $\left\{ \mathrm{A}_{\mu}\right\} $ are now defined as

\begin{align}
\left\{ \mathbf{\mathrm{B}}_{\mu}\right\} =\left\{ \textrm{Ø},\mathrm{\,\overrightarrow{B}}\right\} \,\,\,\,\, and & \,\,\,\,\,\left\{ \mathbf{\mathrm{A}}_{\mu}\right\} =\left\{ \phi,\,\mathrm{\overrightarrow{A}}\right\} .\label{eq:2}\end{align}
Here $\textrm{Ø}$ and $\phi$ are respectively the temporal components
of gravitational and electromagnetic four- potentials, whereas $\overrightarrow{B}$
and $\mathrm{\overrightarrow{A}}$ are the spatial components of the
respective four potentials. Thus, the generalized gravi-electromagnetic
field tensor can be expressed as 

\begin{align}
\mathsf{G_{\mu\nu}=} & \mathsf{F}_{\mu\nu}+i\,\mathsf{f}_{\mu\nu}\label{eq:3}\end{align}
where the real part ( i.e. $\mathsf{F}_{\mu\nu}$) is associated with
linear gravitation and the imaginary part (i.e.$\mathsf{f}_{\mu\nu}$)
is described for electromagnetism. The gravitational $\mathsf{F}_{\mu\nu}$
and electromagnetic $\mathsf{f}_{\mu\nu}$ field tensors are defined
as

\begin{align}
\mathsf{F}_{\mu\nu}= & \mathbf{\mathrm{B}}_{\mu,\nu}-\mathbf{\mathrm{B}}_{\nu,\mu};\nonumber \\
\mathsf{f}_{\mu\nu}= & \mathbf{\mathrm{A}}_{\mu,\nu}-\mathbf{\mathrm{A}}_{\nu,\mu}.\label{eq:4}\end{align}
From equation (\ref{eq:3}), we get

\begin{align}
\mathsf{G}_{\mu\nu}^{\dagger}= & \mathsf{G}_{\mu\nu}^{\dagger}.\label{eq:5}\end{align}
Let us define the gravitational charge $g_{1}$ and electromagnetic
charge $g_{2}$as 

\begin{align}
g_{1}= & \sqrt{k\, m}\,\,\,\,\,\, and\,\,\,\,\, g_{2}=e\label{eq:6}\end{align}
where $k$ is gravitational constant (for brevity we take $k=1$)
and $e$ is electronic charge. As such, we may write the generalized
gravi-electromagnetic charge as complex quantity with gravitational
and electronic charges as its real and imaginary constituents i.e.

\begin{align}
q= & g_{1}+\, i\, g_{2}.\label{eq:7}\end{align}
The generalized current is described as the product of generalized
charge $q$ and four-velocity $\left\{ v_{\mu}\right\} $. So, the
generalized current is expressed as complex quantity as

\begin{align}
\left\{ \mathsf{J}_{\mu}\right\} = & q\,\left\{ v_{\mu}\right\} =\left\{ s_{\mu}\right\} +\, i\left\{ \, j_{\mu}\right\} \label{eq:8}\end{align}
where $\left\{ s_{\mu}\right\} _{\mu}$and $\left\{ \, j_{\mu}\right\} $are
respectively the gravitational and electromagnetic four-currents which
can be obtained from their respective field tensorial equations as 

\begin{align}
\mathsf{F}_{\mu\nu,\nu}= & s_{\mu};\nonumber \\
\mathsf{f}_{\mu\nu,\nu}= & j_{\mu.}\label{eq:9}\end{align}
So, the generalized field equation for gravi-electromagnetism is described
as 

\begin{align}
\mathsf{G_{\mu\nu,\nu}=} & \mathsf{J}_{\mu}\label{eq:10}\end{align}
The Lagrangian density can now be expressed in terms of complex field
tensor, four- potential and four-current as 

\begin{align}
\mathcal{L=}-\frac{1}{4} & \mathsf{G}_{\mu\nu}^{\dagger}\mathsf{G}_{\mu\nu}+\mathsf{V}_{\mu}^{\dagger}\mathsf{J}_{\mu}\nonumber \\
= & -\frac{1}{4}\mathsf{F}_{\mu\nu}\mathsf{F^{\mu\nu}-\frac{1}{4}\mathsf{f}_{\mu\nu}\mathsf{f^{\mu\nu}+A_{\mu}}}j^{\mu}+\mathsf{B_{\mu}}s^{\mu}.\label{eq:11}\end{align}
This Lagrangian density yields the field equation (\ref{eq:9}) followed
by the equation of continuity 

\begin{align}
\partial_{\mu}\mathsf{j^{\mu}}= & \mathsf{j_{\mu,\mu}=}0.\label{eq:12}\end{align}
Hence, the Lorentz force equation of motion is described in the following
form as

\begin{align}
m\frac{d^{2}x_{\mu}}{d\tau^{2}}= & Real(q\,\mathsf{G_{\mu\nu})}v^{\nu}=\left(g_{1}\mathsf{F}_{\mu\nu}+g_{2}\mathsf{f}_{\mu\nu}\right)v^{\nu}.\label{eq:13}\end{align}
Accordingly, the energy momentum tensor may be expressed as 

\begin{align}
\mathsf{T_{\mu\nu}=\mathsf{G_{\mu}^{\sigma}}\mathsf{G_{\nu\sigma}}} & +\frac{1}{4}g_{\mu\nu}\mathsf{G_{\alpha\beta}}\mathsf{G^{\alpha\beta}}\label{eq:14}\end{align}
the components of which are described as 

\begin{align}
\mathsf{T_{00}}= & \frac{1}{2}\left[E^{2}+M^{2}\right]+\frac{1}{2}\left[G^{2}+H^{2}\right]\label{eq:15}\end{align}
and 

\begin{align}
\mathsf{T_{0a}=} & \left[E_{a}+M_{a}\right]+\left[G_{a}+H_{a}\right]\,\,\,(\forall a=1,2,3)\label{eq:16}\end{align}
where $E_{a}$ and $M_{a}$ denote the components of electric and
magnetic fields while $G_{a}$ and $H_{a}$ describe corresponding
components of gravitational and Heavisidian fields.

\section{Quaternion Formalism for Gravi-electromagnetism}

In order to write the quaternion formalism for gravi-electromagnetism
discussed above, let us start with quaternion preliminaries.

\subsection{Quaternion Preliminaries }

The algebra $\mathbb{H}$ of quaternion is a four-dimensional algebra
over the field of real numbers $\mathbb{R}$ and a quaternion $\phi$
is expressed in terms of its four base elements as

\begin{equation}
\phi=\phi_{\mu}e_{\mu}=\phi_{0}+e_{1}\phi_{1}+e_{2}\phi_{2}+e_{3}\phi_{3}(\mu=0,1,2,3),\label{eq:17}\end{equation}
where $\phi_{0}$,$\phi_{1}$,$\phi_{2}$,$\phi_{3}$ are the real
quartets of a quaternion and $e_{0},e_{1},e_{2},e_{3}$ are called
quaternion units and satisfies the following relations,

\begin{eqnarray}
e_{0}^{2} & =e_{0}= & 1\,\,,e_{j}^{2}=-e_{0},\nonumber \\
e_{0}e_{i}=e_{i}e_{0} & = & e_{i}(i=1,2,3)\,\,,\nonumber \\
e_{i}e_{j} & = & -\delta_{ij}+\varepsilon_{ijk}e_{k}(\forall\, i,j,k=1,2,3),\label{eq:18}\end{eqnarray}
where $\delta_{ij}$ is the delta symbol and $\varepsilon_{ijk}$
is the Levi Civita three index symbol having value $(\varepsilon_{ijk}=+1)$
for cyclic permutation , $(\varepsilon_{ijk}=-1)$ for anti cyclic
permutation and $(\varepsilon_{ijk}=0)$ for any two repeated indices.
Addition and multiplication are defined by the usual distribution
law $(e_{j}e_{k})e_{l}=e_{j}(e_{k}e_{l})$ along with the multiplication
rules given by equation (\ref{eq:18}). $\mathbb{H}$ is an associative
but non commutative algebra. If $\phi_{0},\phi_{1},\phi_{2},\phi_{3}$
are taken as complex quantities, the quaternion is said to be a bi-
quaternion. Alternatively, a quaternion is defined as a two dimensional
algebra over the field of complex numbers $\mathbb{C}$. We thus have
$\phi=\upsilon+e_{2}\omega(\upsilon,\omega\in\mathbb{C})$ and $\upsilon=\phi_{0}+e_{1}\phi_{1}$
, $\omega=\phi_{2}-e_{1}\phi_{3}$ with the basic multiplication law
changes to $\upsilon e_{2}=-e_{2}\bar{\upsilon}$.The quaternion conjugate
$\overline{\phi}$ is defined as 

\begin{equation}
\overline{\phi}=\phi_{\mu}\bar{e_{\mu}}=\phi_{0}-e_{1}\phi_{1}-e_{2}\phi_{2}-e_{3}\phi_{3}.\label{eq:19}\end{equation}
In practice $\phi$ is often represented as a $2\times2$ matrix $\phi=\phi_{0}-i\,\vec{\sigma}\cdot\vec{\phi}$
where $e_{0}=I,\,\, e_{j}=-i\,\sigma_{j}(j=1,2,3)$ and $\sigma_{j}$are
the usual Pauli spin matrices. Then $\overline{\phi}=\sigma_{2}\phi^{T}\sigma_{2}$
with $\phi^{T}$ the trans pose of $\phi$. The real part of the quaternion
$\phi_{0}$ is also defined as

\begin{eqnarray}
Re\,\phi & = & \frac{1}{2}(\overline{\phi}+\phi)\,\,,\label{eq:20}\end{eqnarray}
where $Re$ denotes the real part and if $Re\,\phi=0$ then we have
$\phi=-\overline{\phi}$ and imaginary $\phi$ is known as pure quaternion
written as

\begin{equation}
\phi=e_{1}\phi_{1}+e_{2}\phi_{2}+e_{3}\phi_{3}\,\,.\label{eq:21}\end{equation}
The norm of a quaternion is expressed as $N(\phi)=\phi\overline{\phi}=\overline{\phi}\phi=\sum_{j=0}^{3}\phi_{j}^{2}$which
is non negative i.e.

\begin{equation}
N(\phi)=\left|\phi\right|=\phi_{0}^{2}+\phi_{1}^{2}+\phi_{2}^{2}+\phi_{3}^{2}=Det.(\phi)\geq0.\label{eq:22}\end{equation}
Since there exists the norm of a quaternion, we have a division i.e.
every $\phi$ has an inverse of a quaternion and is described as

\begin{equation}
\phi^{-1}=\frac{\overline{\phi}}{\left|\phi\right|}\,\,\,.\label{eq:23}\end{equation}
While the quaternion conjugation satisfies the following property

\begin{equation}
\overline{\phi_{1}\phi_{2}}=\overline{\phi_{2}}\,\overline{\phi_{1}}\,\,\,.\label{eq:24}\end{equation}
The norm of the quaternion (\ref{eq:22}) is positive definite and
enjoys the composition law

\begin{equation}
N(\phi_{1}\phi_{2})=N(\phi_{1})N(\phi_{2})\,\,\,.\label{eq:25}\end{equation}
Quaternion (\ref{eq:17}) is also written as $\phi=(\phi_{0},\vec{\phi})$
where $\vec{\phi}=e_{1}\phi_{1}+e_{2}\phi_{2}+e_{3}\phi_{3}$ is its
vector part and $\phi_{0}$ is its scalar part. So, the sum and product
of two quaternions are described as

\begin{eqnarray}
(\alpha_{0}\vec{,\,\alpha})+(\beta_{0}\vec{,\,\beta}) & = & (\alpha_{0}+\beta_{0},\,\vec{\alpha}+\vec{\beta})\,\,\,,\nonumber \\
(\alpha_{0}\vec{,\,\alpha})\cdot(\beta_{0}\vec{,\,\beta}) & = & (\alpha_{0}\beta_{0}-\overrightarrow{\alpha}\cdot\overrightarrow{\beta}\,,\alpha_{0}\overrightarrow{\beta}+\beta_{0}\overrightarrow{\alpha}+\overrightarrow{\alpha}\times\overrightarrow{\beta})\,\,.\label{eq:26}\end{eqnarray}
Quaternion elements are non-Abelian in nature and thus represent a
non commutative division ring.

\subsection{Quaternion Gravi-electromagnetic Fields}

Let us use the following definitions of electric, magnetic, gravitational
and Heavisidian fields \cite{key-6,key-7} as 

\begin{align}
\overrightarrow{E}= & -\frac{\partial\overrightarrow{A}}{\partial t}-grad\,\phi;\nonumber \\
\,\overrightarrow{M}= & \overrightarrow{\nabla}\times\overrightarrow{A}=curl\,\overrightarrow{A};\nonumber \\
\overrightarrow{G}= & \frac{\partial\overrightarrow{B}}{\partial t}+grad\,\textrm{Ø};\nonumber \\
\,\overrightarrow{H}= & \overrightarrow{\nabla}\times\overrightarrow{B}=curl\,\overrightarrow{B};\label{eq:27}\end{align}
where we use the system of natural units $c=\hbar=1$ throughout the
text along with the gravitational and other constants are being taken
to be unity. As such, we may express the complex four-potential \textbf{$\left\{ \mathsf{V}_{\mu}\right\} $}
in terms of compact quaternion notation as

\begin{align}
\mathbb{V}=\mathsf{V}_{\mu}e_{\mu}= & V_{0}+\sigma_{1}V_{1}+\sigma_{2}V_{2}+\sigma_{3}V_{3}\,\,\,\,\,\,(\mu=0,1,2,3)\label{eq:28}\end{align}
where $\sigma_{1}$, $\sigma_{2}$ and $\sigma_{3}$ are the Pauli
spin matrices which are related to quaternion units as $e_{k}=-\, i\,\sigma_{k}\,\,(\forall k=1,2,3).$The
quaternion differential operator $\mathbb{D}$ may now be expressed
as

\begin{align}
\mathbb{D}= & \partial_{0}+\sigma_{1}\partial_{1}+\sigma_{2}\partial_{2}+\sigma_{3}\partial_{3}\,\,\,\,\,\,(\mu=0,1,2,3).\label{eq:29}\end{align}
Operating the differential operator (\ref{eq:29}) to the quaternionic
four-potential (\ref{eq:28}), we get \cite{key-12},

\begin{align}
\mathbb{D\mathbb{V}} & =\Psi\label{eq:30}\end{align}
where

\begin{align}
\Psi= & \Psi_{0}+\sigma_{1}\Psi_{1}+\sigma_{2}\Psi_{2}+\sigma_{3}\Psi_{3}\label{eq:31}\end{align}
with

\begin{align}
\Psi_{0}=\partial_{0}V_{0}+\overrightarrow{\nabla}\cdot\overrightarrow{V} & =0\label{eq:32}\end{align}
due to Lorentz gauge condition and 

\begin{align}
\Psi_{a}= & \partial_{0}V_{a}+\partial_{a}V_{0}+i\,\left(\overrightarrow{\nabla}\times\overrightarrow{V}\right)_{a}\,\,\,\,(a=1,2,3).\label{eq:33}\end{align}
Using equations (\ref{eq:33}) and (\ref{eq:27}), we get 

\begin{align}
\overrightarrow{\Psi}= & \partial_{0}\overrightarrow{V}+\overrightarrow{\nabla}V_{0}+i\,\left(\overrightarrow{\nabla}\times\overrightarrow{V}\right)\nonumber \\
= & \left(\overrightarrow{G}-\overrightarrow{M}\right)+i\,\left(\,\overrightarrow{H}-\overrightarrow{E\,}\right).\label{eq:34}\end{align}
Thus the generalized electromagnetic field vector $\overrightarrow{\Psi}$
is also expressed as complex quantity like the generalized potential
and current. So, the gravi-electromagnetic field equations (i.e the
generalized Maxwell's Dirac equations) may now be expressed \cite{key-12}
as 

\begin{align}
\overrightarrow{\nabla}\cdot\overrightarrow{\Psi} & =-\mathsf{J_{0};}\nonumber \\
\overrightarrow{\nabla}\times\overrightarrow{\Psi} & =i\,\mathsf{\overrightarrow{J}}-i\,\frac{\partial\overrightarrow{\Psi}}{\partial t}\label{eq:35}\end{align}
Accordingly, we may write the conjugate (i.e. $\overline{\mathbb{D}}$)
of quaternion differential operator $\mathbb{D}$ (\ref{eq:29}) as 

\begin{align}
\overline{\mathbb{D}}= & \partial_{0}-\sigma_{1}\partial_{1}-\sigma_{2}\partial_{2}-\sigma_{3}\partial_{3}\,\,\,\,\,\,(\mu=0,1,2,3).\label{eq:36}\end{align}
Operating (\ref{eq:36}) to (\ref{eq:34}) and using equations (\ref{eq:18}),
we get

\begin{align}
\overline{\mathbb{D}}\,\Psi & =\mathbb{J}\label{eq:37}\end{align}
where

\begin{align}
\mathbb{J}= & J_{0}+\sigma_{1}J_{1}+\sigma_{2}J_{2}+\sigma_{3}J_{3}\label{eq:38}\end{align}
describes the quaternionic form of generalized current (\ref{eq:35})
for gravi-electromagnetism. Accordingly, operating (\ref{eq:29})
to (\ref{eq:38}) and using equations (\ref{eq:18}), we get

\begin{align}
\mathbb{D}\,\mathbb{J} & =\,\mathbb{S}\label{eq:39}\end{align}
where

\begin{align}
\mathbb{S}= & S_{0}+\sigma_{1}S_{1}+\sigma_{2}S_{2}+\sigma_{3}S_{3}\label{eq:40}\end{align}
with 

\begin{align}
S_{0}=\partial_{0}J_{0}+\overrightarrow{\nabla}\cdot\overrightarrow{J} & =0\label{eq:41}\end{align}
due to the equation of continuity and 

\begin{align}
\overrightarrow{S}= & \partial_{0}\overrightarrow{J}+\overrightarrow{\nabla}J_{0}+i\,\left(\overrightarrow{\nabla}\times\overrightarrow{J}\right).\label{eq:42}\end{align}
As such, we may express the generalized gravi-electric $\mathfrak{\overrightarrow{\mathcal{E}}}$
and gravi-magnetic $\overrightarrow{\mathcal{H}}$ fields in terms
of two four-potentials namely the gravi-electric $\left\{ \boldsymbol{B_{\mu}}\right\} $and
gravi-magnetic $\left\{ \boldsymbol{A_{\mu}}\right\} $ as 

\begin{align}
\mathfrak{\overrightarrow{\mathcal{E}}}= & \frac{\partial\overrightarrow{\boldsymbol{B}}}{\partial t}+\overrightarrow{\nabla}\,\textrm{Ø}\textrm{-\ensuremath{\overrightarrow{\nabla}}}\times\boldsymbol{\overrightarrow{A}};\nonumber \\
\mathfrak{\overrightarrow{\mathcal{H}}}=- & \frac{\partial\overrightarrow{\boldsymbol{A}}}{\partial t}-\overrightarrow{\nabla}\,\Phi\textrm{+\ensuremath{\overrightarrow{\nabla}}}\times\boldsymbol{\overrightarrow{B}}.\label{eq:43}\end{align}
These generalized gravi-electric $\mathfrak{\overrightarrow{\mathcal{E}}}$
and gravi-magnetic $\overrightarrow{\mathcal{H}}$ fields have the
similar expressions discussed earlier for the generalized electromagnetic
fields of dyons \cite{key-7,key-10,key-11}in terms of two four-potentials.
As such, the theory of generalized gravito- electromagnetism discussed
here works in the same footing of dyons where the real part is gravitational
sector and imaginary part as electromagnetism leading to striking
symmetry between linear gravitation and electromagnetism.Here dyon
has been considered as the particles carrying simultaneously the existence
of gravitational mass(charge) and the electronic charge. Equation
(\ref{eq:43}) thus satisfies the following differential equations
like the generalized Dirac-Maxwell's (GDM) of dyons \cite{key-7,key-10,key-11}
i.e.

\begin{align}
\overrightarrow{\nabla}\cdot\overrightarrow{\mathcal{E}}= & -\rho_{G};\nonumber \\
\overrightarrow{\nabla}\cdot\overrightarrow{\mathcal{H}}= & \rho_{\mathcal{E}}\nonumber \\
\overrightarrow{\nabla}\times\overrightarrow{\mathcal{E}}= & \frac{\partial\overrightarrow{\mathbb{\mathcal{H}}}}{\partial t}+\overrightarrow{j_{\mathcal{E}}};\nonumber \\
\nabla\times\overrightarrow{\mathcal{H}}=- & \overrightarrow{j_{G}}-\frac{\partial\overrightarrow{\mathcal{E}}}{\partial t};\label{eq:44}\end{align}
where $\rho_{G}$ ; $\rho_{\mathcal{E}}$ are charge source densities
and $\overrightarrow{j_{G}}$; $\overrightarrow{j_{\mathcal{E}}}$
are current source densities respectively associated with gravitational
and electronic charged particles. These GDM type equations (\ref{eq:44})
are invariant under the duality transformations

\begin{align}
\overrightarrow{\mathcal{E}}\longmapsto & \overrightarrow{\mathcal{E}}\,\cos\theta+\overrightarrow{\mathcal{H}}\sin\theta;\nonumber \\
\overrightarrow{\mathcal{\mathcal{H}}}\longmapsto- & \overrightarrow{\mathcal{E}}\,\sin\theta+\overrightarrow{\mathcal{H}}\cos\theta;\nonumber \\
\rho_{G}\longmapsto & \rho_{G}\cos\theta+\rho_{\mathcal{E}}\sin\theta;\nonumber \\
\rho_{\mathcal{E}}\longmapsto- & \rho_{G}\,\sin\theta+\rho_{\mathcal{E}}\cos\theta;\nonumber \\
\overrightarrow{j_{G}}\longmapsto & \overrightarrow{j_{G}}\,\cos\theta+\overrightarrow{j_{\mathcal{E}}}\,\sin\theta\nonumber \\
\overrightarrow{j_{\mathcal{E}}}\longmapsto- & \overrightarrow{j_{\mathcal{E}}}\,\sin\theta+\overrightarrow{j_{G}}\cos\theta.\label{eq:45}\end{align}
On the other hand, the quaternion field equations (\ref{eq:30}),
(\ref{eq:37}) and (\ref{eq:39}) are considered as self-dual. These
are also invariant under quaternion transformations. As such, our
theory is compact, simpler, consistent and manifestly covariant one.
Our theory also reduces to the theory of linear gravity (electromagnetism)
in the absence of electromagnetism (gravitation) or vice versa. It
leads to the dynamics of gravitational (electric) mass (charge) in
the absence of electric (gravitational) charge (mass) or vice versa.
Here dyon has been considered as the  combination of electric charge
and gravitational mass.

\end{document}